\documentclass{elsart}

\usepackage{psfig}

\usepackage{natbib}

\begin{document}

\runauthor{Cicero, Caesar and Vergil}


\begin{frontmatter}

\title{Optical Monitoring of NLS1s at the Wise Observatory}

\author{Ohad Shemmer \and Hagai Netzer}
\address{School of Physics \and Astronomy \and the Wise Observatory, Tel-Aviv
University, Tel-Aviv 69978, Israel}

\begin{abstract}

We report the preliminary results of a spectrophotometric monitoring
program of Narrow-Line Seyfert 1 galaxies conducted at the Wise
Observatory in 1999. The data we collected will be used to evaluate the
line-to-continuum time lag in an attempt to test the idea that the
accretion rate in NLS1s is larger than in other Seyfert 1 galaxies.

\end{abstract}


\end{frontmatter}


\section{Observations and Preliminary Results}

A spectrophotometric monitoring program of 5 Narrow-Line Seyfert 1
(NLS1) galaxies was conducted at the Wise Observatory (WO) during the
summer and fall of 1999.  The observations were performed with the WO
spectrograph, and on separate occasions a CCD camera was used for
broad-band photometry. Long-slit spectroscopy was carried out using a
10" wide slit for both the target NLS1s and nearby comparison stars,
which were verified as non-variables by the photometric monitoring.

\begin{figure}[htb]
\centerline{\psfig{figure=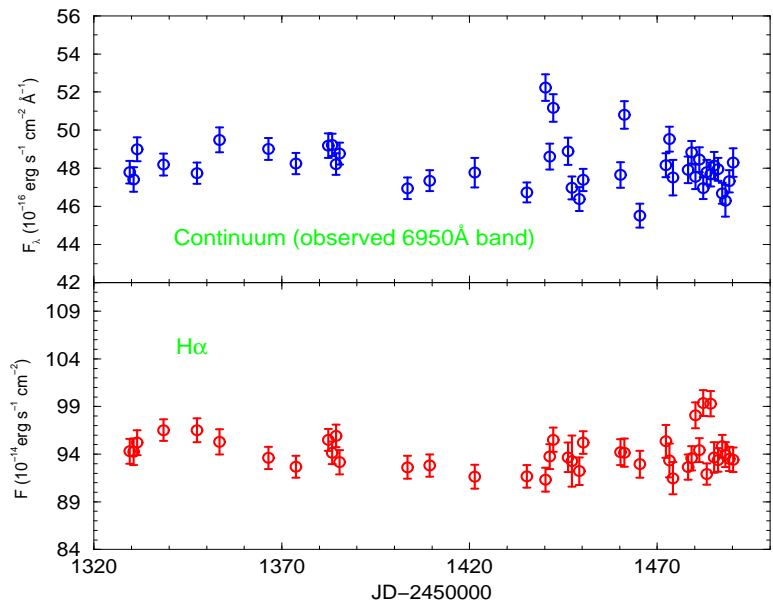,height=3.2truein,width=4truein,angle=0}}
\caption{Akn 564. Light curves obtained from WO observations.}
\label{shemmer_fig_1}
\end{figure}

The prime target of observation was Akn 564, which was under a
continuous monitoring campaign, coordinated by the AGN watch
consortium, from both ground-based observatories and RXTE. In
Figure~\ref{shemmer_fig_1} we present H$\alpha$ and optical continuum
(observed 6950\AA \ band) light curves for Akn 564, observed from WO.
From the light curve it is evident that no significant variability was
recorded for the object either on long or short time scales.   In view
of the extreme X-ray variability of NLS1s (preliminary results from the
RXTE program show that during the first three months of the optical
program, the object varied by over 100\% in X-ray flux), this behavior
is definitely surprising and is atypical of Seyfert 1 galaxies with
broader emission lines.

\begin{figure}[b]
\centerline{\psfig{figure=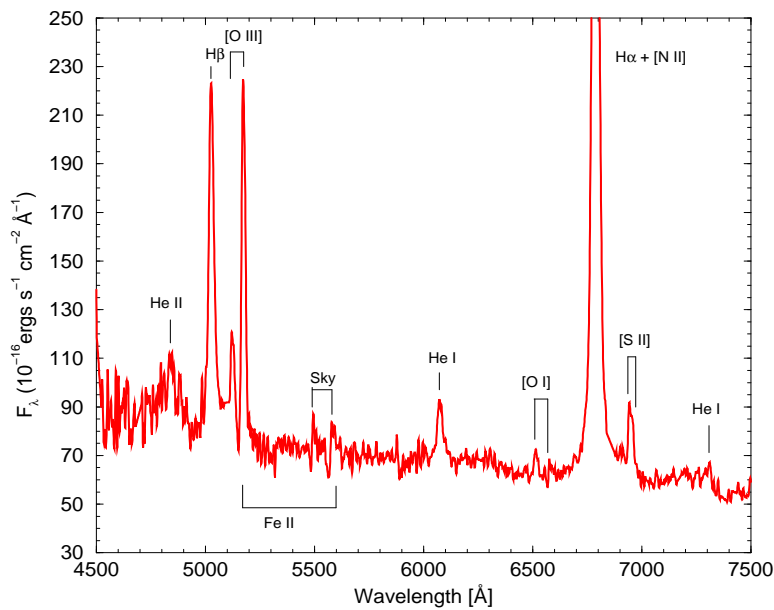,height=3.2truein,width=4truein,angle=0}}
\caption{The spectrum of UCM 2257+2438 obtained at the WO. Note the weak Fe II lines, unlike most NLS1s.}
\label{shemmer_fig_2}
\end{figure}

Two more NLS1s from the sample did not show any significant
variability: I Zw 1 and UCM 2257+2438, the latter (in sharp contrast
with the former) showing very weak Fe II bands as opposed to other
NLS1s (Figure~\ref{shemmer_fig_2}). Mrk 335 and NGC 4051, which were also
in our sample, varied on short as well as on long time scales, as
already seen in previous monitoring campaigns.

\end{document}